\documentclass[12pt]{article}
\usepackage{a4}
\usepackage{epsf}
\newcommand{\beq}{\begin{equation}}
\newcommand{\eeq}{\end{equation}}
\newcommand{\bea}{\begin{eqnarray}}
\newcommand{\eea}{\end{eqnarray}}

\begin{document}

\begin{center}
\title{The fading of symmetry non-restoration at finite temperature}
\author{ M.B. Gavela$^a$, O. P\`ene$^b$, N. Rius$^c$ and S. 
Vargas-Castrill\'on$^a$ } \par \maketitle
$^a$Departamento de F\'\i sica Te\'orica, Universidad Aut\'onoma
de Madrid, Canto Blanco, 28049 Madrid.\\
{$^b$ Laboratoire de Physique Th\'eorique et Hautes
Energies\footnote{Laboratoire
associ\'e au
Centre National de la Recherche Scientifique - URA D00063
\\e-mail: 
gavela@delta.ft.uam.es,
nuria@goya.ific.uv.es,
pene@qcd.th.u-psud.fr,

vargas@delta.ft.uam.es.}}\\
{Universit\'e de Paris XI, B\^atiment 211, 91405 Orsay Cedex,
France}\\
$^c$Departamento de F\'\i sica Te\'orica and IFIC, Centro Mixto 
Universidad de Valencia-CSIC, Valencia, Spain.

\end{center}

\begin{abstract}

 The fate of symmetries at high temperature determines the dynamics
of the very early universe. It is conceivable that temperature effects
favor symmetry breaking instead of restoration. Concerning global
symmetries, the non-linear sigma model is analyzed in detail.
For spontaneously broken gauge symmetries, 
we propose the gauge boson magnetic mass as a 
``flag'' for symmetry (non)-restoration. We consider several cases: 
the standard model with one and two Higgs doublets in the perturbative 
regime, and the case of a strongly interacting Higgs sector.
The latter is done in a model independent 
way with the tools provided by chiral Lagrangians.
Our results clearly point towards restoration, a pattern consistent
with recent lattice computations for global symmetries.
In addition, we explicitly verify $BRST$ invariance for gauge theories 
at finite temperature.
\end{abstract}
\begin{flushright} LPTHE Orsay-98/02\\ FTUAM/97/20\\ 
FTUV/97-73 \\ IFIC/97-105 \\ hep-ph/9801244
\end{flushright}
\newpage

  In which sense does one say that an internal symmetry is restored or broken 
due to temperature effects? What is the relevant order parameter?
And whenever more than one such parameter can be defined, for which physical
consequences are their differences relevant? These are the type of questions 
to face when discussing symmetry (non-)restoration\footnote{ A related
question is the so-called inverse symmetry breaking, describing systems for
which the symmetry is exact at zero temperature and broken when heated; all
through the paper we will take the liberty of dubbing symmetry 
non-restoration both scenarios, unless the contrary is explicitly stated.}. 

 The vacuum structure of a system remains unchanged when it is heated. 
In this sense the degree of symmetry of a system is not modified.
 ``Symmetry restoration'' due to temperature effects is thus a misleading 
denomination for a very simple effect: the spontaneous breaking of 
a global or gauge symmetry can be masked for all physical purposes when 
thermal agitation is present.  This suits intuition,
as a thermal excitation gives in general a positive energy contribution,
allowing particles to ``climb'' barriers between separate minima and 
finally
hiding those barriers for high enough temperatures. Thermal field theory 
computes these effects and usually synthesizes them in the form of a 
so-called
effective potential whose minimum sits at zero values of the fields. 
 Ferromagnets provide well-known experimental examples of a similar behavior
when heated above some critical temperature.

 The suggestion that spontaneously broken field theories are restored at 
high temperature was first made by Kirzhnits and Linde \cite{Linde}. They 
gave qualitative arguments to support this idea in the case of global 
symmetries. In the same direction pointed the results of Dolan and Jackiw 
\cite{DJ} and Weinberg \cite{Weinberg} for gauge theories 
(although in this case the choice of the scalar field vacuum expectation 
value as an order parameter is a delicate one).

  Weinberg noticed as well an opposite possibility: global symmetry 
non-res\-to\-ra\-tion at high temperatures for scalar potentials with more 
than one Higgs multiplet. With just one Higgs the scenario is ruled out 
due to the constraints imposed on the scalar self-coupling 
by the boundedness of the potential, while models with two (or more) 
multiplets can easily accommodate it. The same behavior was found in the
Schwinger model and in a dynamical model of symmetry violation
 in four dimensions \cite{DJ}.

  An analogous situation has been experimentally observed
in nature for the ferroelectric material known as Rochelle salt, which 
shifts from a disordered phase to a more ordered one when heated, as 
measured by the  spontaneous polarization parameter.
In the case of the Rochelle salt the symmetry is restored again for high
enough temperatures, though. Common sense suggests that this should be 
as well the case in field theory, with thermal 
excitations dominating the free energy  unless some finite parameter, 
such as finite volume,
causal domain size, etc., plays a role. 
 Without entering to discuss it, it is clear that even a temporal 
intermediate period, in which thermal effects enhance the effective symmetry
breaking instead of restoring it, could have far reaching
cosmological consequences.

It is worth to remark, though, that Weinberg results on symmetry 
non-restoration are based
on the one-loop approximation to the finite temperature effective
potential, which is known to be unreliable for the discussion
of many aspects of phase transitions. Different techniques, including
non-perturbative ones, are being actively applied to improve the
one-loop approximation, mainly for the study of global symmetries.
The results are very interesting and quite often contradictory:
some studies confirm that symmetry non-restoration exists, although
with a sizable reduction of the parameter space where it occurs
\cite{fight1}-\cite{fight4}, while other analysis conclude that symmetry
is always restored at high temperature when non-perturbative effects are 
taken into account \cite{fight5}-\cite{fight6}.
It has been shown that in a finite lattice no order is possible at
sufficiently high temperature \cite{fight7}. Although the
relevance of this result for the continuum limit is unclear,
a Monte Carlo simulation in 2+1 dimensions seems to support
this conclusion \cite{fight8}.

 Symmetry non-restoration is indeed being increasingly reconsidered as
a candidate way out of many cosmological problems arising in spontaneously 
broken theories. Examples are the domain wall and axion problems \cite{axion}
and the monopole problem in Grand Unified Theories \cite{monopole}.

 As recalled in Section \ref{gauge}, in the minimal standard 
$SU(2)\otimes U(1)$ model the symmetry is necessarily restored, given the 
simplicity of its Higgs sector. 
 At present, there are two main avenues to explore physics beyond the 
Standard Model: theories in which the Higgs particle is a fundamental one, 
supersymmetry being its most representative example, and those for which 
it is not, currently dubbed as strongly interacting Higgs scenarios.

 Supersymmetry is broken ``de facto'' at high temperatures, due to the 
difference
in the boson and fermion populations, as dictated by Bose-Einstein versus 
Fermi-Dirac
 statistics. The debatable and interesting question is whether the 
internal symmetries present in supersymmetric theories, and whose fate is
fundamental for the existence of topological defects, are restored. 
It has been proven that such is the case for 
renormalizable supersymmetric theories \cite{haber}.
  For the latter, a recent
analysis for systems involving non-vanishing background charges shows that 
symmetry non-restoration could be possible \cite{central}.
 The consideration of non-renormalizable terms in the Lagrangians has
 led as well to a polemics: their mere addition does not lead to 
symmetry non-restoration \cite{senja}. 
   
 Here we rather follow the path leading to a non-elementary Higgs scenario.
In so doing, we first reanalyze the global $SU(N_f)_R \otimes SU(N_f)_L$
non-linear sigma model, relevant in supergravity and many other 
scenarios, in sect. \ref{sigma}. 
Sect. \ref{gauge} is devoted to the analysis of gauge symmetries;
after discussing  $BRST$ invariance at finite temperature, 
we study the behavior of the $SU(2) \otimes U(1)$ symmetry in several 
scenarios. In subsect. \ref{perthiggs} we analyze both the minimal 
standard model and the standard model with two Higgs doublets
within the perturbative regime, while 
in subsect. \ref{strong} we consider a strongly interacting Higgs 
sector in a model independent way, using the techniques
of chiral Lagrangians, and we discuss the differences with the results of the
previous section. These different chapters are preceded by some comments 
on order parameters, sect.\ref{order}, and followed by our conclusions.

\section{The order parameter}
\label{order}
The interesting order parameter to consider in a phase transition 
depends first of all on the question one wants to study.

An illustrative example is provided by spin systems in solid state physics.
Both in ferromagnets and antiferromagnets, the ground state breaks 
rotational symmetry: the spins align for the former and display an
antiparallel alignment  for the 
latter. The traditional order parameter is the average spontaneous 
magnetization $\langle \vec{m} \rangle \neq 0$ which plays a 
crucial role
in the description of the response of the system to an external magnetic 
field: it turns to be important for ferromagnets, while marginal
 or even vanishing for antiferromagnets to the extent 
that the ground state approaches the N\'eel-type magnetic order. Hence, the
spontaneous magnetization is an example of order parameter whose non-zero 
value is not necessary for the spontaneous breakdown of the symmetry. 
 Ferrimagnets are yet another scenario: antialignment is present alike to
the case of antiferromagnets although $\langle \vec{m} \rangle \neq 0$, as the 
weight allocated to the two possible spin projections differs.

  Analogous questions arise in particle physics: different so-called order 
 parameters  can be  correlated to different physical effects.
The appropriate parameter depends on the aspect of the history
 of the universe under study, and not all of them necessarily ``bip''
 simultaneously.  

  Already at zero temperature, the relationship among different
possible order parameters is not always straightforward. Recall 
massless $QCD$ at low energies, with pion interactions appropriately
described by chiral Lagrangians. The pion decay constant, $F_\pi$, 
and the condensate, $\langle \bar\Psi \Psi \rangle$, are not 
necessarily equivalent order parameters. Although unnatural,
  $\langle \bar\Psi \Psi \rangle = 0$ is not theoretically forbidden
 while a non-null v.e.v. of some higher dimension operator 
accompanies $F_\pi$ as a ``flag'' for dynamical symmetry breaking \cite{stern}.

  In a general way it is clear that when the Lagrangian, at zero temperature,
 is just a one parameter theory, all putative order parameters should be 
 equivalent.  Such is the case with most Lagrangians respecting  global 
 symmetries, where the value of the field at the minimum of the effective 
 potential is a well-defined 
 order parameter, commonly used, and  any other one is simply related to 
 it.

  On the contrary, for spontaneously broken gauge theories
  the issue is much more subtle. To begin 
 with, the v.e.v. of any non-gauge invariant operator is necessarily
 zero \cite{elitzur}. 
 Only gauge invariant operators, such as $|\phi|^2$, may have a 
 non-vanishing v.e.v., signaling the Higgs mechanism. 
 Once a gauge-fixing procedure has been performed, a gauge-non invariant
 minimum of the effective potential may appear, which may be useful whenever
 its physical meaning is properly extracted in due respect of general Ward
 identities. The same applies to the v.e.v. of higher dimension operators.

  The effective potential itself is gauge dependent. However, the values of 
the effective potential at its local minima or maxima are gauge independent.
  Hence if there is a minimum of $V(\phi)$ with a value lower than $V(0)$
in one gauge, then there will be such minimum in any gauge (although its
position will generally be different) and the symmetries will definitely
be broken.

 In practice, gauge-dependent correlation functions
  are often used in the study of phase transitions: the physical conclusions
  are expected to be 
 rather close to those derived with gauge-invariant ones if the fluctuations 
 of the scalar fields are small compared to their vacuum expectation 
 values.
 This can be safe in the broken phase of the theory, while quite misleading
 in the symmetric phase, as recently discussed in ref. \cite{buch}, where a 
 detailed description of the zoo of correlation functions can be found as well.

  It is worth to briefly specify the ``flags'' for symmetry breaking discussed
 in the present paper:

 - for the global symmetries of the non-linear $\sigma$ model and its 
extensions in terms of chiral Lagrangians, we discuss both
 the pion decay constant $F_{\pi}$ and the vacuum expectation value of the
 condensate. The latter is  defined from an effective potential and can be 
interpreted as the  remnant of the 
 disappeared sigma field. Both parameters are essentially equivalent 
since, at zero
temperature, we are dealing with a one parameter theory. 

 - for the gauge symmetry case, specifically the standard electroweak model
 and its extensions,
 we concentrate instead on particle masses.
 In the perturbative regime, 
 both the negative
 Higgs ``mass''  and the magnetic mass for the gauge bosons are discussed. 
When the Higgs particle disappears from the spectrum and we enter the 
 non-perturbative regime of the gauged non-linear sigma model, our order 
parameter will be the gauge boson magnetic mass.

 The magnetic mass squared is defined as the temperature dependent
contribution to the transverse part of the gauge boson self-energy, 
$\Pi_T (0,\vec{k})$, for vanishing three-momentum $\vec{k}$. 
At the order we work it is gauge invariant. Notice that  
Weinberg \cite{Weinberg} advocates the use of gauge invariant operators 
carrying moderate momenta and zero energy as order parameters.

The analogous electric mass, whose square is defined by the longitudinal
 component of the gauge boson self-energy,
$\Pi_{L} (0,\vec{k})$ with $\vec{k}\rightarrow {0}$, is not a 
suitable parameter. Indeed, it tends to increase at high temperature
even when the symmetry is restored. The intuitive explanation is electric 
screening: some particles in the theory carry an electric charge. 
 Already at one-loop order, thermal fluctuations pull 
charged 
pairs out of the vacuum to screen external charges. However, there are no 
fundamental particles in any gauge theory which carry a magnetic charge. 
Magnetic screening can presumably then only arise from non-perturbative 
fluctuations which carry magnetic charge.

 A perturbative computation of the magnetic mass shows that it is exactly 
 equal to zero at one loop in an 
unbroken gauge theory. For unbroken non-Abelian gauge theories, such as QCD,
  higher orders in perturbation theory suffer from
infrared divergences, and a magnetic mass of 
order $g^2 T$ is expected to be generated 
non-perturbatively. In spontaneously broken gauge theories, such as the
 standard 
electroweak model and its extensions, no such divergences are present. 
Thus, we propose to use the magnetic mass as a ``flag'', expecting 
that even in perturbation theory it will show 
a tendency to vanish at high enough temperatures 
whenever symmetry restoration occurs. Of course, it will be a valid ``flag''
only when exploring the broken phase of the theory, for the reasons given
above.

 Another pertinent point to recall is that temperature corrections break 
Lorentz invariance as the plasma sets a preferred reference frame.
Assume for instance a zero temperature Lagrangian based on an internal 
symmetry.
Certainly the mixed states describing the new ``effective vacuum'' may 
greatly differ from the real vacuum structure. 
What about the finite temperature effective Lagrangian itself? 
Up to which point its functional form may differ from the initial one?
Non-zero temperature is tantamount to treat space and time differently: 
internal symmetries at the Lagrangian level, such as chiral symmetry, 
cannot be explicitly broken due to it.
What is to be a priori expected is a splitting of any operator into 
 its temporal 
and spatial components, with differing coefficients. 
For instance $F_{\pi}$ in the chiral non-linear sigma model will generate 
two different coupling constants at finite temperature, 
a temporal one, $F_{\pi}^t(T)$, and a spatial one, $F_{\pi}^s(T)$
\cite{ptyt}. We leave the corresponding considerations for gauge theories 
for the beginning of sect. \ref{gauge}.  
 
A necessary condition for symmetry restoration is that all possible order 
parameters or ``flags'' for symmetry restoration do signal it.

\section{Global non-linear sigma model: the $T \neq 0$ effective Lagrangian}
\label{sigma}

 The restoration of spontaneously broken global symmetries is discussed 
in this section within an effective Lagrangian approach.
We consider the $SU(N_f)_R \otimes SU(N_f)_L$ non-linear 
sigma model, which may be defined by the Lagrangian
\begin{equation}
{\cal L} = \frac{1}{2} 
\partial_{\mu} \pi_a \partial^{\mu} \pi_a +  \frac{1}{2}
\partial_{\mu} \sigma \partial^{\mu} \sigma \, ,
\end{equation}
with the constraint 
\begin{equation}
F_\pi^2 = \sigma^2  + \vec{\pi}^2   \ .
\label{cons} 
\end{equation}
By convention, we take the scalar condensate in the direction
of the $\sigma$ component; that is, at tree level 
\begin{equation}
\langle  \sigma^2  \rangle =  F_\pi^2 \ .
\end{equation}
Since the global symmetry is broken down to an $SU(N_f)$ symmetry, there
are  a total of $N_f^2-1$ Goldstone bosons, which are identified with the 
pion fields, $\vec {\pi}$. 
The constraint (\ref{cons}) determines the $\sigma$ field in terms 
of the pion fields, so that in the non-linear Lagrangian only the 
latter appears.
A non-linear redefinition of the fields is possible without 
changing the physical content of the theory, leading to 
different parametrizations. The so-called exponential representation 
can be described by the Lagrangian 
\begin{equation}
{\cal L}^{(2)} = \frac{1}{4} ~ F_{\pi}^2 ~
Tr (\partial_{\mu} U \partial^{\mu} U^\dagger),
\label{chirlag}
\end{equation}
where $U$ is a $SU(N_f)$ unitary matrix field
\begin{equation}
U = \exp \left( i \, \frac{\pi_a T_a} {F_{\pi}} \right), 
\label{exp}
\end{equation}
with $T_a$ the generators of $SU(N_f)$, normalized as 
${\rm Tr} (T_a T_b) = 2 \delta_{ab}$ and
$[T_a,T_b] = 2 i f_{abc} T_c$, being $f_{abc}$ the structure constants 
of $SU(N_f)$.

As it is well known, all realizations of the non-linear chiral 
Lagrangian, such as the exponential one (\ref{chirlag}),
square root, Weinberg, etc. \cite{Kapusta}, with $N_f = 2 (3)$, 
are low energy effective theories for QCD with 2(3) massless quarks,
expressed in terms of Goldstone bosons and systematically expanded in 
powers of the Goldstone bosons momenta.
As a consequence of the chiral symmetry, these models possess the 
remarkable property of {\it universality}: 
once the coupling constants have been adjusted ($F_\pi$
is the only one at lowest order) all physical predictions are the same.
Therefore, the chiral Lagrangian not only
parametrizes the dynamics of the Goldstone bosons that emerge in QCD but 
also of any other theory, such as the Higgs model, that follows 
the same symmetry breaking pattern.

We have analyzed two order parameters: the pion decay constant $F_\pi$, 
and the vacuum expectation value of the $\sigma$ field,
$\langle  \sigma  \rangle$. Notice that, while their behavior should be
 essentially equivalent, their precise variation rate with temperature may 
differ somewhat. Indeed, the 
constraint (\ref{cons}) as a thermal average implies 
$\langle  \sigma^2  \rangle =  F_\pi^2$, while in general 
$\langle  \sigma^2  \rangle \neq \langle  \sigma  \rangle^2$.
Our treatment differs from previous ones in that we have considered them
as Lagrangian parameters, whose variation with temperature is read from 
the one-loop effective Lagrangian we derive.

Although we will just discuss below the calculation in the exponential 
representation, we have explicitly checked that the results of measurable 
quantities are the same in other parametrizations used in the literature, 
namely the square root and Weinberg representations.
Of course, for quantities without a physical meaning, the temperature 
corrections can be representation dependent.

We drop all temperature independent ultraviolet divergent quantities from our
expressions, recalling that when a theory is renormalized at zero 
temperature no more infinities of that type appear at finite temperature.

\subsection{Temperature corrections to $F_{\pi}$}

Temperature corrections to $F_{\pi}$ are obtained from an effective 
Lagrangian approach. 
The chiral Lagrangian (\ref{chirlag}) can be expanded in powers of 
$(\frac{\pi}{F_{\pi}})^2$ up to a certain order, 

\begin{equation}
{\cal L}^{(2)} = 
\frac{1}{2} \partial_{\mu} \vec{\pi} \partial^{\mu} \vec{\pi} + 
\frac{1}{6 F^{2}_{\pi}} ((\vec{\pi} \partial_{\mu} \vec{\pi}) (\vec{\pi} 
\partial^{\mu} \vec{\pi}) - (\vec{\pi}  \vec{\pi}) \partial_{\mu} 
\vec{\pi} \partial^{\mu} \vec{\pi}) + \ldots
\end{equation}

We have computed the one-loop temperature corrections  
to this Lagrangian to leading order $T^2 / F^{2}_{\pi}$, and 
proved that they lead to an effective Lagrangian 
with the same structure as the tree-level one, albeit with
two $F_{\pi}$'s: a temporal one, $F_{\pi}^t$, and a spatial one,
$F_{\pi}^s$.
It encloses the full temperature effects
in the renormalized (temperature dependent) parameters.

Due to the derivative character of the interactions, a 
contribution to the kinetic energy term, at leading order
 $T^2 / F^{2}_{\pi}$, is 
obtained when computing the one particle irreducible (1PI) two 
point Green function at one loop (Fig. 1).
This term is absorbed by pion field renormalization. In the exponential
representation used here, we find
\beq
\pi^2 (T) = \pi^2 
\left[ 1 - \frac{(N - 1) T^2}{36 F^{2}_{\pi}} \right].
\label{piones}
\eeq

\begin{figure}
\begin{center}
\makebox[3cm]{
\epsfysize 3cm
\epsfbox[149 266 305 345]{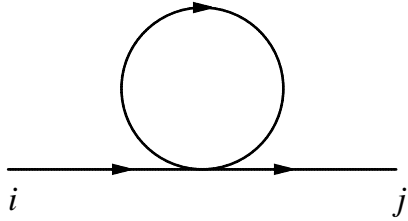}
}
\caption{One loop self-energy diagram for the pions.} 
\end{center} \end{figure}

Diagrams in Fig. 2 contribute to the 1PI four point function at one 
loop, leading to different thermal corrections for the spatial and the 
temporal coupling constants in which $F_\pi$ splits at finite 
temperature, as mentioned above.
To this order, it results
\bea
F^{s}_{\pi} (T) &=& F_{\pi} 
\left[ 1 - \frac{(N - 1) T^2}{24 F^{2}_{\pi}} \right],
\\
F^{t}_{\pi} (T) &=& F_{\pi} \left[1 - \frac{(N + 1) T^2}{24 F_{\pi}^{2}}
\right],
\label{fpi}
\eea
where $N= N_f^2 -1$ represents the number of pions. 
Both temperature dependent renormalized parameters 
$F^s_{\pi}(T)$ and $F^t_{\pi}(T)$ show a 
clear tendency to vanish at high enough temperatures, pointing towards 
chiral symmetry restoration. We have as well explicitly checked that 
$F^s_{\pi}$ so derived is representation independent.

Thermal corrections to the pion decay constant have been 
computed in the literature following different 
approaches \cite{Leut1}, \cite{Schenk}, \cite{Kapusta}, 
\cite{ptyt2}, \cite{Cris}.
Our result for the effective spatial coupling, $F^s_{\pi}(T)$, is 
in agreement with those calculations of $F_{\pi}(T)$. 
In most of them, $F_{\pi} (T)$ is obtained from its usual definition
(slightly modified at finite temperature \cite{Kapusta})
through the two point function of the axial vector current, 
and there is no splitting between temporal and spatial couplings
at one loop; it appears at two loops \cite{toublan}.
Notice that since we consider $F_\pi$ just as a parameter in the 
Lagrangian, it does not necessarily coincide with the 
pion decay constant as usually defined.

\begin{figure}
\begin{center}
\leavevmode
\epsfysize=3.0truecm
\epsffile{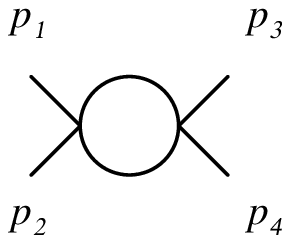}
\epsfysize=3.0truecm
\epsffile{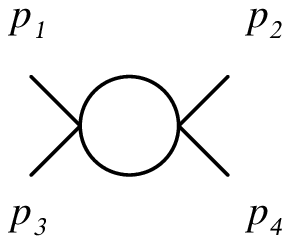}
\epsfysize=3.0truecm
\epsffile{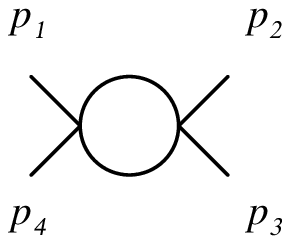}
\\
\vspace*{0.5cm}
\epsfysize=3.0truecm
\epsffile{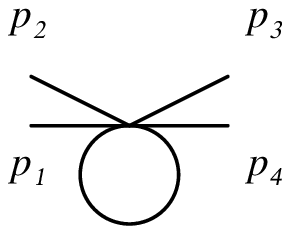}
\caption{One loop diagrams contributing to the 1PI four point 
Green function for the pions.}
\end{center} 
\end{figure}

To avoid technical complications, we have computed the pion field and 
$F_{\pi}$ 
renormalization from the lowest order terms in the expansion of the 
Lagrangian ${\cal L}^{(2)}$ in powers of the pion fields; 
chiral symmetry ensures that all higher terms in 
the field expansion are consistently renormalized once $\vec{\pi}$ and 
$F_{\pi}$ have been renormalized from these lowest order terms.

\subsection{Temperature Corrections to the Condensate at One Loop}

Temperature corrections to $\langle \sigma \rangle$ 
are computed through the addition of a small chirality breaking term, 
which makes the Lagrangian slightly asymmetric
\footnote {We recall that the QCD scalar density  $\overline{\Psi} \Psi$
whose vacuum expectation value represents the 
familiar QCD condensate, is 
equivalent to $\sigma \equiv \frac {F_\pi} {4} Tr (U + U^+)$ 
since $\overline{\Psi} \Psi$ and $Tr (U + U^+)$ can be shown to 
transform in the same way under the chiral group.
${\cal L}_B$ is thus equivalent to a quark mass term.},
that is, we consider 
\beq
{\cal L} = {\cal L}^{(2)} + {\cal L}_B ,
\eeq
with
\begin{equation}
{\cal L}_B = c \sigma = \frac{c F_{\pi}}{4} ~ Tr(U + U^\dagger).
\label{cb}
\end{equation}
Expanding the last term in powers of $\frac{\pi^2}{F^2_{\pi}}$ in 
the exponential representation it is found
\begin{equation}
{\cal L}_{B} = c F_{\pi} - \frac{c}{2 F_{\pi}} \vec{\pi} \vec{\pi} + 
\frac{c}{24 F^{3}_{\pi}} (\vec{\pi} \vec{\pi})^2 + \ldots
\label{ay}
\end{equation}
Following the effective Lagrangian approach, we compute the 
one-loop order $T^2$ corrections to ${\cal L}$
through the 1PI zero, two and four point Green functions.
The kind of diagrams involved in the calculation are vacuum energy ones for 
the zero point 1PI Green function and the 
same as in the previous section (see Figs. 1 and 2), 
although with modified couplings, for the two and four point 1PI functions.  
Now, besides $F_\pi$ and the pion field, also the 
parameter $c$ is renormalized. Since it only appears 
in the product $c F_\pi$, there is an ambiguity, depending 
on which (spatial or temporal) $F_\pi(T)$ we consider, 
leading to 

\beq
c^s(T)=c \left( 1 - \frac{T^2}{24 F_\pi^2} \right),
\eeq

\beq
c^t(T)=c \left( 1 + \frac{T^2}{24 F_\pi^2} \right).
\eeq
Again, the one-loop effective Lagrangian written in terms of the 
temperature-dependent parameters has the same structure as 
the tree-level one, space-time splitted, though.

Notice that (minus) the first term in the expansion of 
${\cal L}_{B}$, $- c F_{\pi}$, 
can be interpreted as the vacuum energy density of the system. That is,
the free energy of a system of free bosons, given by 
\beq
c^s (T) \,  F^s_{\pi}(T) = c^t(T) \, F^t_{\pi}(T) \, .
\eeq

Since the operator $\sigma$ can be obtained by deriving the bare Lagrangian 
with respect to the parameter $c$ (see eq. (\ref{cb})), we can also 
interpret the result as a thermal correction to the scalar condensate
\footnote{Recall that the thermal average of the operator 
$\sigma$ is defined as 
\beq
\langle \sigma \rangle_T = \frac 
{Tr (\sigma e^{-\beta H})}{Tr (e^{-\beta H})} 
\eeq  
where $Tr e^{-\beta H} = \int [dU] e^{-\int d^4 x {\cal L}}$ is the
partition function and $\beta = 1/T$. Thus, one can compute 
$\langle \sigma \rangle_T$ as the derivative of the 
partition function with respect to the parameter $c$, at $c=0$.
This is analogous to the extraction of 
$\langle \bar{\Psi} \Psi \rangle_T$ in QCD, by
first adding an explicitly chiral symmetry breaking term $m \bar{\Psi} \Psi$
 to the bare Lagrangian, computing the temperature corrections, and deriving
 then with respect to $m$ \cite{Leut1}.}.
Taking the derivative with respect to (bare) $c$ of the one loop 
effective Lagrangian, the temperature corrections to the condensate 
are found. Explicit chiral 
symmetry is recovered by fixing $c = 0$ at the end of the computation.
The final result is satisfactorily the same whether
either the spatial set $F_{\pi}^s, c^s$ or the temporal set $F_{\pi}^t , 
c^t$ is used, leading to:   

\begin{equation}
\langle \sigma \rangle_T = 
\langle \sigma \rangle \left(1 - N \frac{T^2}{24 F_{\pi}^{2}} \right),
\label{con}
\end{equation}
in agreement with \cite{Gaillard}, \cite{Leut1}, \cite{Kapusta}.
As can be seen in (\ref{con}), the temperature correction to the condensate 
also points towards chiral symmetry restoration.

Notice that ${\cal L}_{B}$ is just the well known classical 
potential up to a minus sign. However, 
it was not possible to use the standard method for computing
effective potentials \cite{DJ} due to the presence of derivative 
couplings. Using a generalization of this method  
\cite{Gaillard} the same result is recovered, as already mentioned.

   
\section{Gauge symmetry: $SU(2)\otimes U(1)$}
\label{gauge}

In this section we study theories with gauge group $SU(2)\otimes U(1)$.
We consider the cases with one light Higgs doublet, two light Higgs 
doublets, and the
generic one where the Higgs sector becomes strongly interacting, the 
latter done in a model independent way. 
Before entering into such details, we dwell again into the delicate
issue of the ``flag'' for
symmetry (non-)restoration for gauge theories, and in the fate of 
gauge invariance itself when a system is heated. 

\subsection{The magnetic mass}

  In a gauge theory, the pseudo-Goldstone bosons
of the Lagrangian  are unphysical fields, unlike the gauge bosons.

  As stated in sect. \ref{order}, we choose the gauge boson magnetic mass
 as our ``flag'' or indicator for symmetry (non-)restoration. 

At non-zero temperature, the self-energy tensor of the gauge boson 
may depend on the four-velocity of the plasma $u_{\mu}$. Consequently, 
the gauge boson self-energy can be expressed as a 
linear combination of four possible tensors: $g_{\mu \nu}$, $k_{\mu} 
k_{\nu}$, $u_{\mu} u_{\nu}$ and $k_{\mu} u_{\nu} + k_{\nu} u_{\mu}$.
Some linear combinations of these tensors
are usually chosen as the standard basis set \cite{basis}, denoted  
$A_{\mu \nu}, B_{\mu \nu}, C_{\mu \nu}$ and $D_{\mu \nu}$ and 
defined in Appendix A. 
In this basis, the one loop gauge boson self-energy is written as
\begin{equation}
\Pi^{\mu \nu}= 
\Pi_{T} A^{\mu \nu} + \Pi_{L} B^{\mu \nu} + \Pi_D D^{\mu \nu} ,
\label{au} 
\end{equation}
where the subscripts $T$ and $L$ denote transverse and longitudinal with 
respect to the spatial component $\vec{k}$ of the wave vector.

The magnetic mass is defined as $\Pi_T (0,\vec{k})$, with vanishing 
$\vec{k}$.
At one-loop and 
leading order, $O(gT)$, it will be shown to be gauge invariant both for 
the standard model and for its extensions considered below. 
The explicit computations will be focused in the W gauge boson mass.

\subsection{Checking gauge invariance: $BRST$ identities}

 Up to our knowledge, the Slavnov-Taylor identities at finite temperature
have never been explicitly verified in the literature for the 
electroweak theory. 
We explicitly perform such a task in the present work, for the 
two-point functions of the theory.

  Indeed, one expects gauge invariance to be preserved at non-zero 
temperature. A simple reasoning can
be developed in the imaginary time formalism, where finite temperature just
amounts to compactifying the time direction, that is, to perform a global 
``distortion'' of the system. Gauge transformations are local ones by 
definition, and thus they
should not be affected by  global topological conditions.
Once the gauge fixing procedure has been implemented, $BRST$ 
invariance remains,
and the corresponding Slavnov-Taylor identities are to be proven.
 
  One should realize that the proof is much more juicy than at zero 
temperature: there, quadratic divergences are disposed of  by counterterms
 from the start, and the Slavnov-Taylor identities for such non-physical 
quadratically divergent terms are not even considered. At finite temperature
 those quadratic divergences are the source of the $T^2$ dependence. It is 
then mandatory, and new, to check the $BRST$ identities on them.

Both in the linear and the non-linear realizations of the 
$SU(2)\otimes U(1)$ gauge symmetry, 
the Ward identities relating the two point Green functions at one loop 
are given by 

\begin{eqnarray}
k^2 (\Pi_{D}^{W} + 2 M_{W} \Pi^{W \pi^{\pm}}) - 
M_{W}^{2} \Pi^{\pi^{\pm}} &=& 0, \nonumber \\
k^2 (\Pi_{D}^{Z} - 2 i M_{Z} \Pi^{Z \pi^{0}}) - M_{Z}^{2}
\Pi^{\pi^{0}} &=& 0, \nonumber \\
k^2 \Pi_{D}^{\gamma} &=& 0, \nonumber \\ 
k^2 (\Pi_{D}^{\gamma Z} - i M_{Z} \Pi^{\gamma \pi^{0}}) &=& 0, 
\nonumber \\ 
\xi \Pi_D^Z - i M_Z \xi \Pi^{Z \pi^0} - \Pi^{c^0} &=& 0, 
\nonumber \\
\xi \Pi_D^W + M_W \xi \Pi^{W \pi^{\pm}} - \Pi^{c^{\pm}} &=& 0,
\label{wi}
\end{eqnarray}
where $\xi$ is the gauge fixing parameter in $R_{\xi}$ gauges.
$\Pi_{D}^{W}$, $\Pi_{D}^{Z}$, $\Pi_{D}^{\gamma}$ and 
$\Pi_{D}^{\gamma Z}$ are the form factors introduced in eq. ({\ref{au})
for the $W^{\pm}-W^{\pm}$, $Z-Z$, $\gamma - \gamma$
and $Z - \gamma$ self-energies, respectively.
$\Pi^{W \pi^{\pm}}$ is defined from the two point Green function 
with external legs 
$W^{\pm}-\pi^{\pm}$ as
\beq
\Pi_\mu^{W \pi^{\pm}} = k_\mu \Pi^{W \pi^{\pm}},
\eeq
and analogously for $\Pi^{Z \pi^{0}}$ and $\Pi^{\gamma \pi^{0}}$,
while $\Pi^{c^{\pm}}$and $\Pi^{c^0}$ 
represent the charged and 
neutral Faddeev-Popov ghosts self-energies.


\subsection{Perturbative Higgs sector}
\label{perthiggs}
\subsubsection{Minimal Standard Model}
\label{standard}

Consider the minimal electroweak standard model, that is, with just one 
light Higgs doublet. Here, all the couplings of the theory are 
in the perturbative range, and we can rely on the one-loop approximation 
to the effective potential at non-zero temperature.

It is well known that with just one Higgs doublet the gauge symmetry is
always restored at high temperature. Indeed, given the simplicity of
the potential,
\begin{equation}
V(\phi)_{T = 0} = - \mu^{2} (\phi^{\dagger} \phi) + \lambda (\phi^{\dagger} 
\phi)^{2}, 
\label{vsm}
\end{equation}
the condition that it has to be bounded from below forces the sign
of $ \lambda $ to be positive. 
The one-loop thermal corrections to the above potential can be
readily computed  in $R_\xi$ gauges by the usual methods \cite{DJ}.
In the high temperature limit ($T \gg m_i$, with $m_i$ the masses of 
all standard model particles) the leading order $T^2$ 
corrections are gauge invariant and read:

\begin{equation}
\delta \mu^2 = - \frac{T^2}{12} (6 \lambda + \frac{9}{4} g^2 + 
\frac{3}{4} g'^{2} + 3 h_{t}^2 + 3 h_{b}^2 + h_{\tau}^2),
\end{equation}
leading to the temperature dependent Higgs v.e.v.
\begin{equation}
v(T)^2 = \,v^2 -\frac{T^2}{2} \left[1+\frac{3 g^2}{8\lambda} 
+ \frac{g'^2}{8\lambda} + \frac{h_{t}^2}{2\lambda} +
 \frac{h_{b}^2}{2\lambda} 
+ \frac{h_{\tau}^2}{6\lambda} \right]
\label{srsm}
\end{equation}
where $v^2= \mu^2/\lambda$ denotes the Higgs v.e.v. at tree level and 
$h_{t}$, $h_{b}$ and $h_{\tau}$ are the Yukawa coupling constants of the 
quarks $t$ and $b$, and the $\tau$ lepton, respectively.

\begin{figure}[h]
\begin{center}
\leavevmode
\epsfysize=2.0truecm
\epsffile{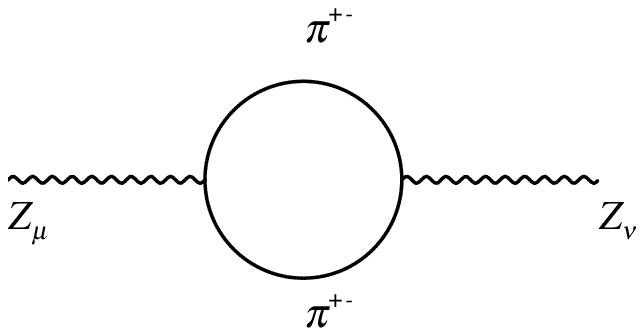}
\epsfysize=2.0truecm
\epsffile{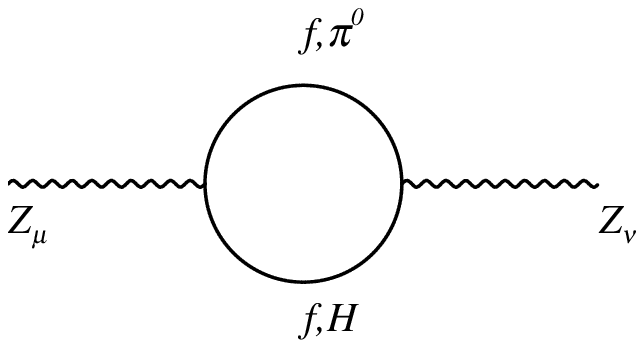}
\epsfysize=2.0truecm
\epsffile{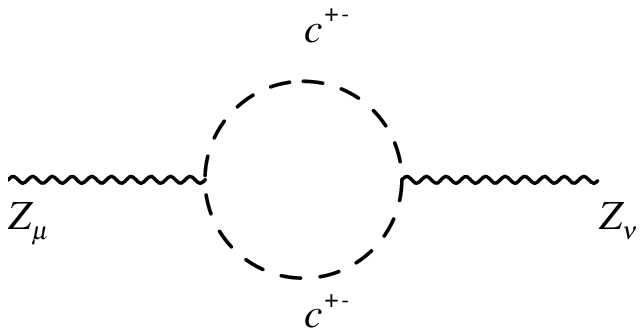}
\epsfysize=2.0truecm
\epsffile{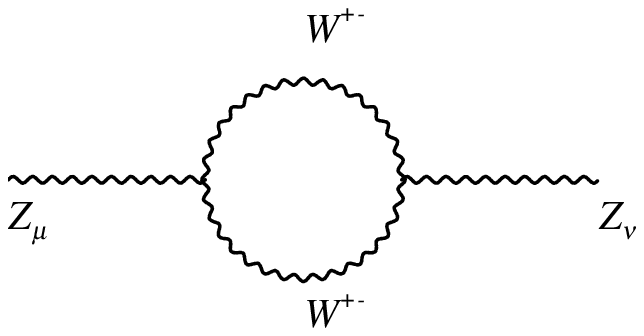}
\\
\vskip 0.3truecm
\leavevmode
\epsfysize=2.0truecm
\epsffile{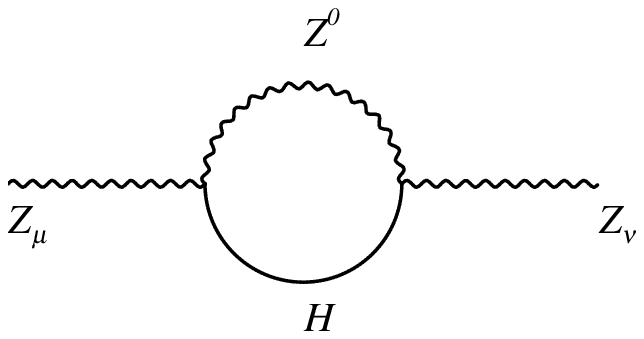}
\epsfysize=2.0truecm
\epsffile{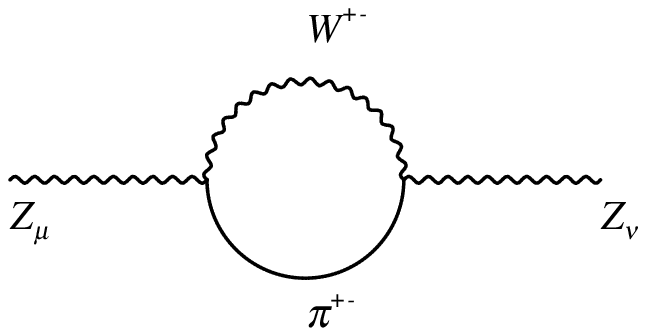}
\epsfysize=2.6truecm
\epsffile{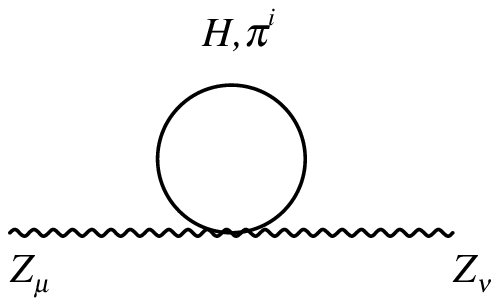}
\epsfysize=2.6truecm
\epsffile{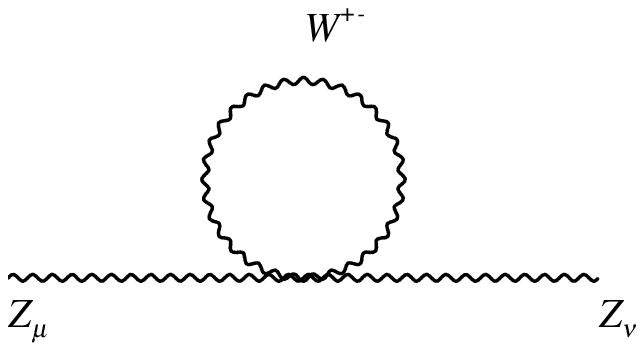}
\\
\vskip 0.3truecm
\leavevmode
\epsfysize=2.5truecm
\epsffile{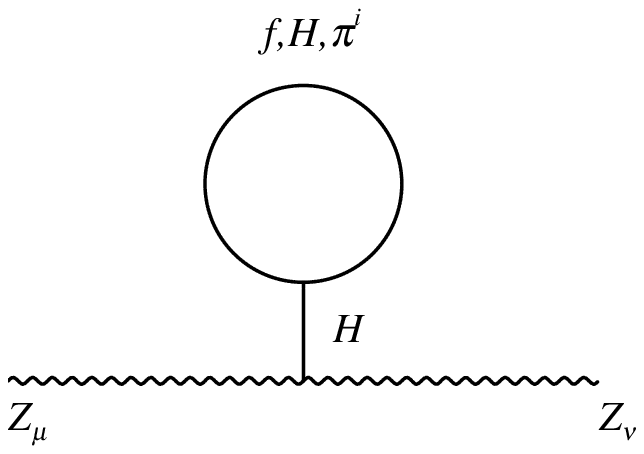}
\epsfysize=2.5truecm
\epsffile{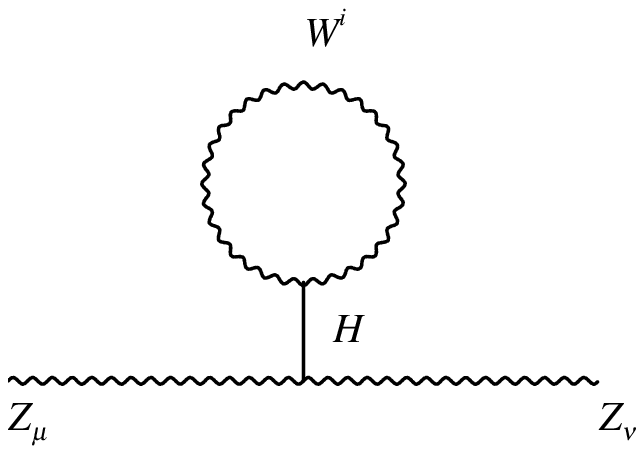}
\epsfysize=2.5truecm
\epsffile{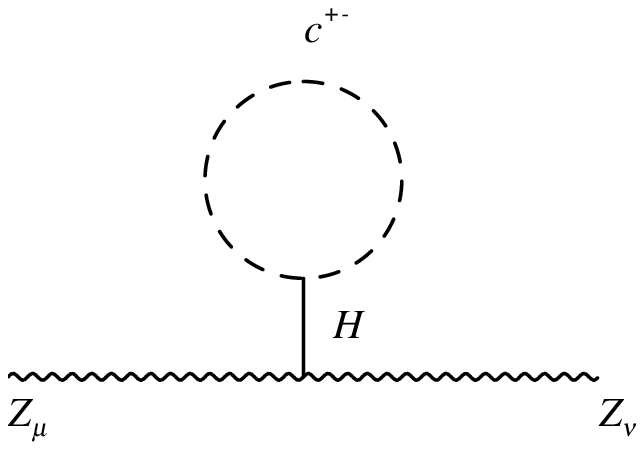}
\\
\caption{One loop self-energy diagrams for the $Z$ gauge boson in the 
minimal standard model.} 
\label{smgb}
\end{center} 
\end{figure}

As expected, exactly the same behavior is seen from the $Z$ and $W$ gauge 
boson magnetic mass.
The set of diagrams contributing to the $Z$ self-energy at one loop are 
shown in Fig. \ref{smgb}. The coupling constants are 
not renormalized at one loop (at order $T^2$), which allows to write 
\begin{eqnarray}
M_{W,mag}^2 &=& \frac{g^2} {4}\, v(T)^2, \\ 
M_{Z,mag}^2 &=& \frac{g^2 + g'^2} {4}\, v(T)^2, \nonumber 
\label{magstand} 
\\
v(T)^2 &=& \,v^2 -\frac{T^2}{2} \left[1+\frac{3 g^2}{8\lambda}
+ \frac{g'^2}{8\lambda} + \frac{h_{t}^2}{2\lambda} +
 \frac{h_{b}^2}{2\lambda} + 
\frac{h_{\tau}^2}{6\lambda} \right] . 
\nonumber
\end{eqnarray}

The result in eq. (\ref{srsm}) is then recovered, 
pointing towards restoration in an inescapable way. 

Finally, we have checked all the Ward identities in eq. (\ref{wi});
the explicit results for the diagrams involved can be found  
in Appendix C.

\subsubsection{Two Higgs doublets}
\label{twohiggs}

Models with a richer Higgs structure have several scalar couplings.
In order to explore symmetry non-restoration, the rule of the game is then 
to play with the freedom in the sign of some of those couplings,
while respecting the boundedness condition.

The simplest extension, i.e., the standard model with two Higgs doublets, 
is considered now.
We make the usual assumption that the down quarks and charged leptons
only couple to the Higgs doublet $\phi_1$ and the up quarks to   
$\phi_2$, ensuring tree-level flavor conservation of scalar
mediated neutral currents. 
In order to avoid radiatively induced FCNC terms, 
we also impose the discrete
symmetry $\phi_1 \rightarrow -\phi_1$.
The most general, renormalizable, scalar potential consistent with the
above symmetry and with gauge invariance is:
\bea
V(\phi_1, \phi_2) &=&
-m_1^2 \phi_1^\dagger \phi_1 -  m_2^2 \phi_2^\dagger \phi_2
+ \lambda_1 (\phi_1^\dagger \phi_1)^2  
+ \lambda_2 (\phi_2^\dagger \phi_2)^2 \nonumber \\
& & + \lambda_3 (\phi_1^\dagger \phi_1) (\phi_2^\dagger \phi_2)  
+ \lambda_4 |\phi_1^\dagger \phi_2|^2
+ \frac{1}{2} [\lambda_5 (\phi_1^\dagger \phi_2)^2 + h.c.].
\label{pot}
\eea
The condition for the potential to be bounded from below
leads to the constraints:
\bea
& & \lambda_1>0 , \ \  \lambda_2>0 , \ \ 
4 \lambda_1 \lambda_2 > \lambda_3^2 , \nonumber \\
& & 4 \lambda_1 \lambda_2 > (\lambda_3 + \lambda_4 + \lambda_5)^2
\ \ \ \ (for ~ \lambda_5 < 0 ).
\label{stab}
\eea

The leading one-loop thermal corrections give the following 
thermal masses for the fields $\phi_1, \phi_2$: 
\bea
\Delta V_{T}(\phi_1, \phi_2) &\simeq&
\frac{T^2}{12}\left[(6\lambda_1 + 2 \lambda_3 + \lambda_4
+ \frac{9}{4} g^2 + \frac{3}{4} g'^2 + 3 h_b^2 + h_\tau^2) |\phi_1|^2
\right. \nonumber \\
& & + \left. (6 \lambda_2 + 2 \lambda_3 + \lambda_4
+ \frac{9}{4} g^2 + \frac{3}{4} g'^2 + 3 h_t^2) |\phi_2|^2 \right]
\nonumber \\
&  \equiv & m_1^2(T) |\phi_1|^2  + m_2^2(T) |\phi_2|^2 .
\label{pert}
\eea
 Although the contributions from both fermions and gauge bosons are
positive, the scalar couplings $\lambda_3$ and $\lambda_4$ may be
negative, and therefore it is not possible to make any
a priori statement about the signs of the mass terms above.
What can be stated is that the stability conditions in eq.(\ref{stab})
do not allow both mass terms in eq. (\ref{pert}) to be negative.
Since $\phi_2$ receives a large positive contribution from the
top Yukawa coupling, it is easier to get a negative thermal mass
for the field $\phi_1$. Then, its vacuum expectation
value would remain non-zero at high temperature and
the $SU(2) \otimes U(1)$ symmetry would never be restored.

\vspace*{1.3cm}

\input epsf

\hbox{
        \vbox{\epsfysize=6.0truecm
        \epsffile{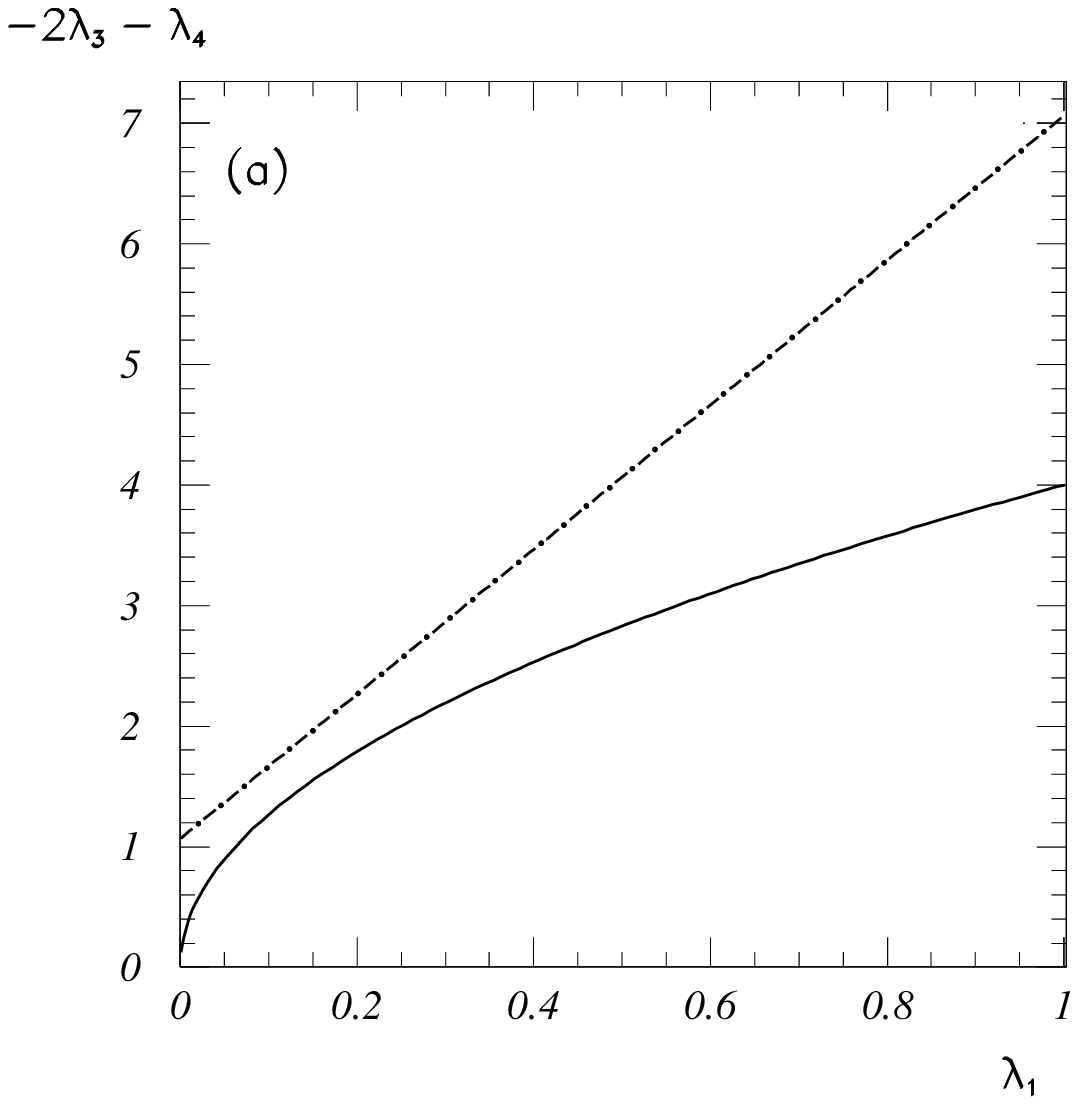} }
\hskip 1truecm
        \vbox{\epsfysize=6.0truecm
        \epsffile{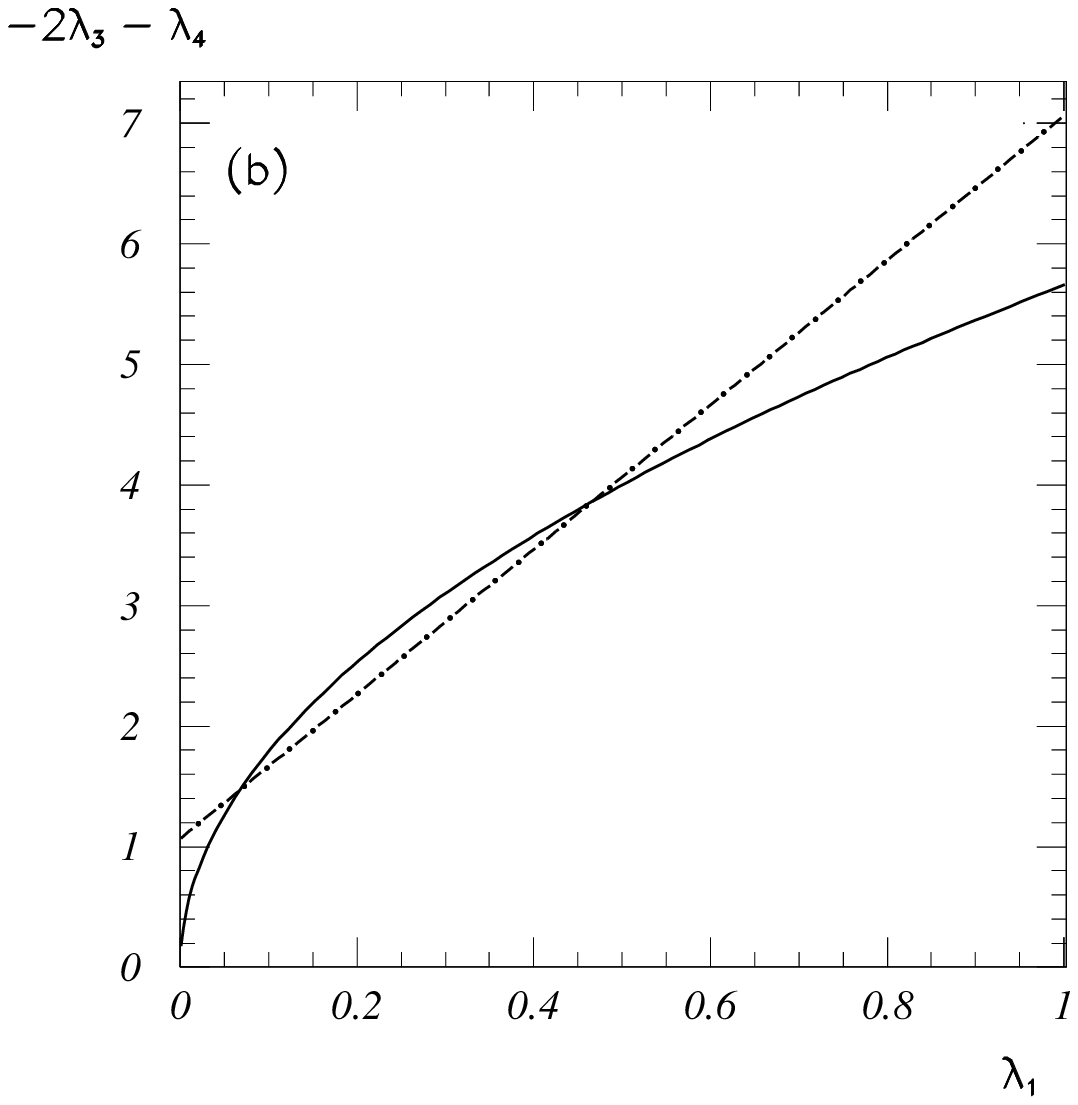} }
}
\begin{figure}[h]
\begin{center}
\vspace*{5mm}
\caption{Stability bound of the tree level potential (solid line) and
parameter space leading to symmetry non-restoration (above the
dashed-dotted line), for $\lambda_2=1$ (a) and $\lambda_2=2$ (b), with
$\lambda_5=0$ in both.}
\end{center}
\end{figure}

According to eq. (\ref{pert}), $m_1^2(T) < 0$ requires
\beq
6\lambda_1 + 2 \lambda_3 + \lambda_4
+ \frac{9}{4} g^2 + \frac{3}{4} g'^2 + 3 h_b^2 + h_\tau^2 < 0 
\label{m1} \,.
\eeq
Notice that the term $\frac{9}{4} g^2 = 0.99$ is already of order one
at the electroweak scale, which makes it difficult to attain
$m_1^2(T) < 0$  within the perturbative regime
\footnote{We neglect small finite temperature renormalization of the
couplings.}. 
We have checked numerically that the condition
(\ref{m1}) and the stability bounds (\ref{stab}) are incompatible
for scalar couplings in the range $[-1 < \lambda_i < 1]$, where the 
weak coupling
expression (\ref{pert}) is justified.
As an example, in Fig. 4 we plot the stability bound of the tree 
level potential (the allowed range is below the solid line) and 
the curve corresponding to $m_1^2(T) = 0$ (symmetry non-restoration
occurs above the dashed-dotted line), for $\lambda_5=0$
and $\lambda_2=1,2$.

Heading outside the above mentioned range, the numerical results taken at
face value seem to 
point towards the possibility of symmetry non-restoration as the scalar
sector enters the non-perturbative regime, as can
be seen in Fig. 4b. Of course, the above computation is meaningless
outside that range.

 On the above, we have used as ``flag'' for symmetry (non-)restoration the
negative scalar ``masses'', that is, the location of the minimum of the
potential. As discussed in previous section, one could calculate instead the 
induced gauge boson magnetic masses, 
as an alternative analysis of the fate of the symmetry. 
In the linear realization of the symmetry breaking sector of the
minimal 
standard model, the magnetic mass has been computed and proved to show 
a tendency towards vanishing. The temperature corrections to the vacuum
expectation value of the Higgs field indirectly computed through this
procedure agree with the result obtained from the effective potential 
approach.
The magnetic mass squared is given by $g^2 \frac{v^2 (T)}{4}$. In
the two doublet case, the magnetic mass squared would be given by 
$g^2 \frac{v_{1}^{2} (T) + v_{2}^{2} (T)}{4}$, thus if there is a region of 
parameter space for which one of the vev's remains non-zero the magnetic 
mass will not show a tendency towards vanishing and symmetry 
non-restoration becomes possible.

 From our study of the two doublet model, we conclude that the 
requirement of 
the validity of perturbation theory points towards the usual assumption of 
restoration of the $SU(2) \otimes U(1)$ gauge symmetry.
In the next section, we extend the analysis outside the perturbative
regime.

\subsection{Strongly Interacting Higgs Sector}
\label{strong}

We now study the behavior of the $SU(2)_L \otimes U(1)_Y$ symmetry in 
the Standard Model with a strongly coupled Higgs sector.
Strong coupling implies (at least naively) heavy physical scalar 
particles, which can be effectively removed from the physical low-energy 
spectrum.
An effective Lagrangian approach is the natural technique to use when 
all the physical degrees of freedom in the symmetry breaking sector are 
heavy. 
We then consider the most general effective Lagrangian 
which employs a non-linear realization of the spontaneously 
broken $SU(2)_L \otimes U(1)_Y$ gauge symmetry \cite{al}.
The resulting chiral Lagrangian is a non-renormalizable non-linear
$\sigma$ model coupled in a gauge invariant way to the Yang-Mills theory.
 Chiral Lagrangians have been widely used in the last few years
as low energy effective theories for electroweak interactions 
\cite{Quim}.

The Lagrangian keeps only the light degrees of freedom, namely the gauge 
and Goldstone bosons. The latter are collected in a unitary matrix
$U = \exp (i \pi_a \tau_a/v)$, where $v$ is the vacuum expectation value 
that gives the $W$ and $Z$ gauge bosons a mass, $\pi_a$ are the would-be
Goldstone fields and $\tau_a$ the Pauli matrices.

\subsubsection{The lowest order Lagrangian}

The lowest order terms in a derivative expansion of the effective
Lagrangian are 
\begin{equation}
{\cal L}_{GChL} = \frac{v^2}{4}{\rm Tr} [D_{\mu}U^{\dag}D^{\mu}U]
+{\cal L}_{YM}+{\cal L}_{GF} +{\cal L}_{FP}, 
\label{gcl}
\end{equation}
where
\begin{equation}
D_{\mu} U=\partial_{\mu}U+i\frac{g}{2}(\vec{W}_{\mu} \vec{\tau}) U
-i \frac{g'}{2} U (B_{\mu} \tau^3)  \ .
\end{equation}
${\cal L}_{YM}$ is the pure Yang-Mills piece 
\begin{equation}
{\cal L}_{YM}= - \frac{1}{2} Tr (W_{\mu\nu} W^{\mu\nu}) -
\frac{1}{4} B_{\mu\nu} B^{\mu\nu} ,
\end{equation}
and we consider the following gauge-fixing term
\begin{equation}
{\cal L}_{GF} = - \frac{1}{2} 
\left( \frac{1}{\sqrt{\xi_1}} \partial_{\mu} W^{\mu}_i 
- g \frac{v}{2} \sqrt{\xi_2} \pi_i \right)^2 -
\frac{1}{2} \left( \frac{1}{\sqrt{\xi_1}} \partial_{\mu} B^{\mu} - 
g' \frac{v}{2} \sqrt{\xi_2} \pi_3 \right)^2 ,
\label{gf}
\end{equation}
from which the Faddeev-Popov term, ${\cal L}_{FP}$, can be computed
in the usual way. The relevant part for our calculation is 
given in Appendix B. 
At tree level we take $\xi_1 = \xi_2$,  
so that the gauge boson - Goldstone boson mixing term is canceled.  

Expanding $U$, we obtain the interaction vertices, in particular the 
tree level masses are given by
\bea
M_Z^2 = (g^2 + g'^2) \, \frac{v^2}{4} & , & 
M_{W}^2 = g^2 \, \frac{v^2}{4} \, ,\\
m_{\pi^0}^2 = \xi_2 M_Z^2  & , & m_{\pi^{\pm}}^2 = \xi_2 M_{W}^2 \, ,
\\
m_{c^0}^2 = \sqrt{\xi_1 \xi_2} \, M_Z^2  & , & 
m_{c^{\pm}}^2 = \sqrt{\xi_1 \xi_2} \, M_{W}^2 \, ,
\label{masstl}
\eea
where $\pi^0, \pi^{\pm}$ are the longitudinal components of the gauge 
bosons $Z, W^{\pm}$, respectively, and $c^0, c^{\pm}$ are the 
corresponding ghost fields.

\begin{figure}[h]
\begin{center}
\leavevmode
\epsfysize=1.4truecm
\epsffile{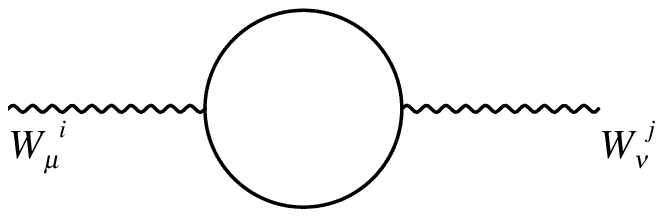}
\vspace*{-0.5truecm}
\epsfysize=2.2truecm
\epsffile{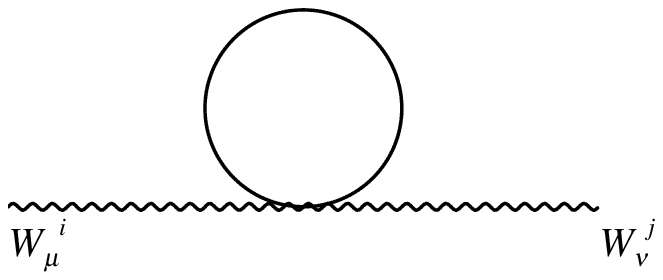}
\\
\vspace*{0.8truecm}
\leavevmode
\epsfysize=1.4truecm
\epsffile{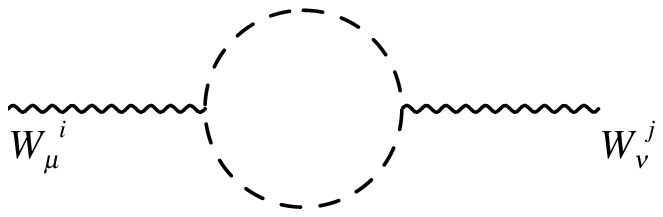}
\epsfysize=1.4truecm
\epsffile{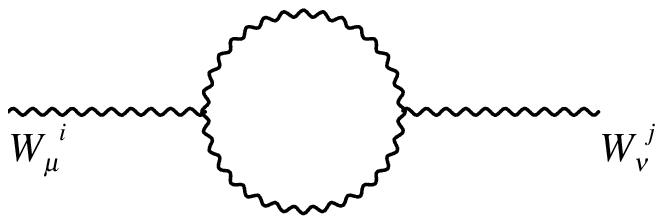}
\vspace*{-0.5truecm}
\epsfysize=2.2truecm
\epsffile{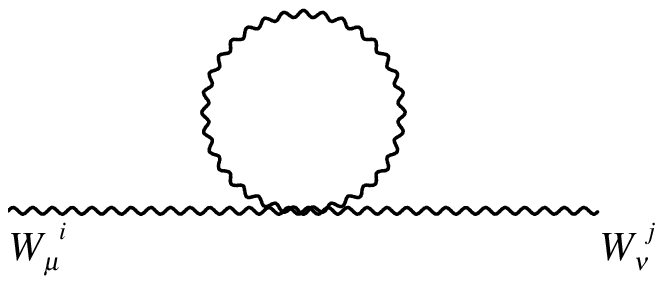}  
\\
\caption{One loop self-energy diagrams for the gauge bosons in the 
standard model with strongly coupled Higgs sector. Solid lines 
represent would-be-Goldstone bosons, wavy lines gauge bosons and 
dashed lines Faddeev-Popov ghosts.} 
\label{gb}
\end{center} 
\end{figure}

We have computed the one loop temperature corrections to the effective 
Lagrangian (\ref{gcl}) at leading order, i.e., ${\cal O} (T^2)$.
Before entering the discussion of the results, it is 
worth to recall the range of validity of the calculation 
and the approximations involved. 
Unitarity implies that this low-energy effective theory 
should be valid for an energy scale much smaller than 
$4 \pi v  \sim 3 $ TeV. Furthermore, as we shall see in eq.(\ref{mmag}),
our approximation cannot be valid unless $T<\sqrt 6\, v$, 
where the latter limit would give the naively
extrapolated critical temperature. In the vicinity of it, the loop 
expansion performed here is not appropriate.    
Therefore, our conclusions will be reliable 
up to $T\simeq 200$ GeV, and expected to be an acceptable guideline
up to 500 GeV. 
In order to obtain analytic expressions
we work in the limit $T \gg m_i$, with $m_i$ the masses of the low 
energy spectrum  (which means $T \gg g v$), and $T \gg k$, where $k$ 
are the external momenta. This approximation is known in the literature 
as the hard thermal loop approximation (HTL) \cite{bp}. 
 
We are doing 
an expansion in both $T / v$ and the small coupling
constants $g, g'$. The corrections of order $g^2 \frac {T^2}{v^2}$
are smaller than $\frac {T^4} {v^4}$, which will only appear  
at higher order in perturbation theory since, in the HTL 
approximation, we assume $g v \ll T$.
Indeed, notice that already at $T=0$  the
gauge coupling constants, $g,g'$, are not corrected  by quadratically 
divergent diagrams, as can be seen from simple power counting arguments
 on the one-loop diagrams contributing to
vertex functions: only logarithmic divergences appear. Hence no $T^2$ 
renormalization of $g, g'$ may appear.

Let's start with the thermal corrections to the gauge boson masses,
obtained from the corresponding self-energy tensor (Fig. \ref{gb}). 
It is well known that the magnetic mass of the gauge bosons 
in an unbroken gauge theory vanishes at one loop \cite{mag}, so 
that only diagrams involving would-be-Goldstone boson loops will give a 
non-zero 
contribution to the magnetic masses in the broken phase, and we find 
\footnote{Although we have not included fermions in our calculation, 
it is easy to see that they do not contribute to the magnetic masses 
either, at leading order.}

\begin{eqnarray}
M_{W,mag}^{2} &=&  g^2 \frac{v^2}{4} 
\left( 1 - \frac{N_f T^2}{12 v^2} \right), 
\nonumber \\
M_{Z,mag}^{2} &=&  (g^2 + g'^2) \frac{v^2}{4} 
\left( 1 - \frac{N_f T^2}{12 v^2} \right), 
\nonumber \\
M_{\gamma,mag}^{2} &=& 0, 
\label{mmag}
\end{eqnarray}
where $N_f=2$ in $SU(2)$. 
For the electric masses, defined as $\Pi_L(0,\vec{k})$,
we get \footnote{We acknowledge C. Manuel for
pointing out two misprints in these formulae, in an earlier version.} 
\bea
M^{2}_{W,el} &=&  g^2 \frac{v^2}{4} 
\left( 1 + \frac{17 N_f T^2}{12 v^2} \right),
\nonumber \\
M^{2}_{Z,el} &=&  (g^2 + g'^2) \frac{v^2}{4} 
\left( 1 - \frac{N_f T^2}{12 v^2} \right) 
+ \frac{N_f T^2}{24}  [(g^2 + g'^2)(c_W^2 - s_W^2)^2 + 8 g^2 c_W^2] 
\nonumber \\
M^{2}_{\gamma,el} &=&  e^2 \, \frac{N_f T^2}{2},
\label{mel}
\eea
where $c_W$ ($s_W$) is the cosine (sine) of the weak mixing angle at 
zero temperature.

Focusing on the magnetic masses, we can then rewrite them as
\bea
M_{W,mag}^{2} &=&  g^2 \, \frac{v(T)^2}{4} \ ,
\\ 
M_{Z,mag}^{2} &=&  (g^2 + g'^2) \, \frac{v(T)^2}{4} \ , 
\eea 
with $v(T)^2$ given by 
\bea
v(T)^2 &=& v^{2}
\left[ 1 - \frac{(N - 1) T^2}{12 v^{2}} \right].\  
\label{vev}
\eea 
where $N=3$ in $SU(2)$.

It is worth to remark that the would-be-Goldstone boson field 
renormalization is the same as the one for the Goldstone bosons in the 
global case (eq. (\ref{piones})), while the temperature corrections 
to $v^2$ coincide with those of ${F_\pi^s}^2$.

The diagram in Fig. \ref{mix} generates a gauge boson - Goldstone boson 
mixing term proportional to $T^2$, which is absorbed by a 
renormalization of the gauge fixing parameter $\xi_2$:
\beq
\xi_2(T) = \xi_2 \left( 1 +  \frac{2 T^2}{9 v^2} \right).
\label{xi2}
\eeq

\begin{figure}[h]
\begin{center}
\makebox[5cm]{
\epsfysize 5.5cm
\epsfbox[142 198 397 397]{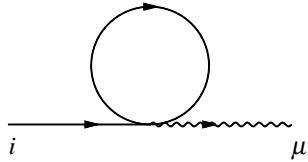}
}
\caption{One loop diagram which generates the $\pi - W$ mixing term.}
\label{mix}
\end{center} \end{figure}

\begin{figure}
\begin{center}
\makebox[2.8cm]{
\epsfysize 2.8cm
\epsfbox[142 260 327 357]{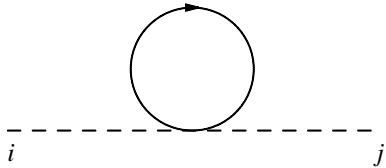}
}
\caption{One loop self-energy diagram for the Faddeev-Popov ghosts.}  
\label{fanta}
\end{center}
\end{figure}

With respect to the remaining parameter in the effective Lagrangian in
 eq. (\ref{gcl}), the gauge fixing one $\xi_1$, it is not renormalized at  
${\cal O} (T^2)$. The same applies to the gauge boson and ghost fields.  
Naive dimensional counting shows that the 1PI one-loop diagrams that 
will renormalize those entities are at most logarithmically divergent,
alike to the situation for $g$ and $g'$ discussed earlier, 
and thus not able to produce $T^2$ corrections.

\begin{figure}[h]
\begin{center}
\leavevmode
\epsfysize=2.2truecm
\epsffile{aupi1.eps}
\hspace*{0.5truecm}
\epsfysize=2.3truecm
\epsffile{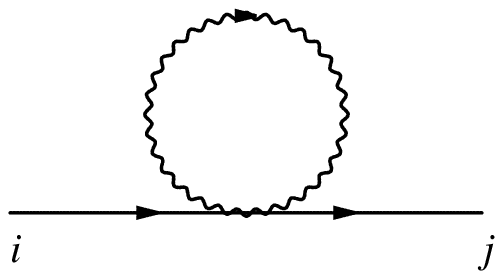}
\\
\vspace*{1.0truecm}
\leavevmode
\epsfysize=2.0truecm
\epsffile{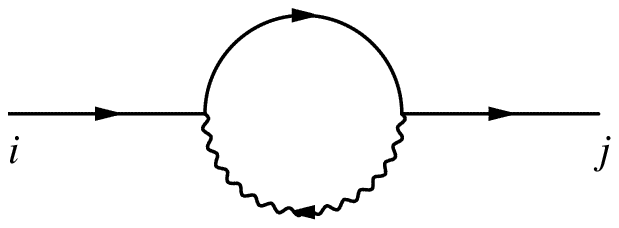}
\hspace*{0.5truecm}
\epsfysize=2.4truecm
\epsffile{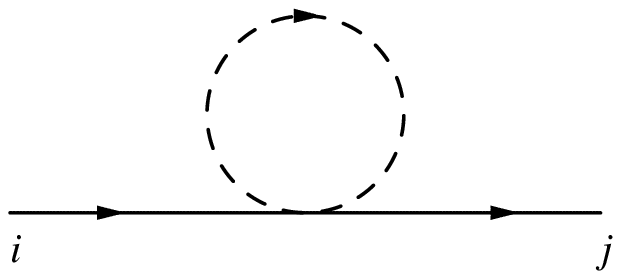}  
\\
\caption{One loop self-energy diagrams for the would-be-Goldstone bosons.}
\label{aupis}
\end{center} \end{figure}

The consistency of our results has been verified by computing the leading 
order corrections to the masses of ghost and 
would-be-Goldstone bosons through the corresponding  
one loop self-energies (Figs. \ref{fanta} and \ref{aupis}). 
Once the would-be-Goldstone boson field 
renormalization has been taken into account, the temperature-dependent 
masses depend on the renormalized parameters in the same way 
as at tree level, i.e.,  
\bea
m_{\pi^{\pm}}^2(T) &=& \xi_2(T) M_{W,mag}^2 ,
\\
m_{c^{\pm}}^2(T) &=& \sqrt{\xi_1 \xi_2(T)} \, M_{W,mag}^2 .
\eea

The one-loop effective Lagrangian does not have exactly
the same functional form as the original bare one, eq. (29): it splits,
as exemplified by the differing electric and magnetic masses. 
Eqs. (40) and (41) allow to connect several important finite $T$ quantities
in a compact notation, though. Eqs. (42) and (43) are an example of it.

We have also checked all the Ward identities in eq. (\ref{wi}).
The explicit results for the different diagrams involved are given 
in Appendix C.

A natural question is the relationship with the linear case discussed in 
subsect. \ref{standard}: one expects 
that taking there the Higgs mass to infinity the results of the present 
section should be recovered.
A superficial look does not show it. For instance, taking the limit
$\lambda \rightarrow \infty$ in eq. (\ref{magstand}) for the magnetic 
mass in the linear case, eq. (\ref{mmag}) is not recovered. 
There is no inconsistency,
though: the high temperature and heavy Higgs mass limits are not 
interchangeable. We have indeed checked that eq. (\ref{mmag}) 
is obtained by taking the limit 
$m \rightarrow \infty$ 
($m$ being the physical Higgs mass), 
before doing the one-loop computation in the linear case (which
implies not considering the following diagrams of Fig. 3: 2, 5, 10, 11 
and 7, 9 when the physical Higgs boson is in the loop).  
The same argumentation is valid for the rest of the physical parameters.

Regarding the question of the $SU(2) \otimes U(1)$ symmetry non-
restoration which initially motivated our study, we conclude 
from eq. (\ref{mmag}) that thermal effects tend to restore the 
symmetry also in the non-perturbative regime. 
Notice that although the magnetic mass is non-vanishing in the 
symmetric phase beyond one-loop order, it is expected to be 
of order $g^2 T$, and therefore much smaller than the 
magnetic mass in the broken phase, of order $g v$
(recall that $g^2 T \ll gT < g v $).
Thus we can interpret the decreasing of the magnetic mass with 
the temperature as a ``flag'' for symmetry restoration.
On the contrary, for instance the $W$
electric mass in eq. (\ref{mel}) can be
written as 
\beq
M_{W,el}^{2} =  g^2 \, \frac{v^2(T)}{4} + g^2 \,\frac{3 N_f T^2}{8}, 
\eeq
and in the symmetric phase, while $v (T)=0$, it is nevertheless 
non-vanishing and of order $gT$, 
as anticipated.

\begin{figure}
\begin{center}
\makebox[8cm]{
\epsfysize 8cm
\epsfbox[7 157 533 652]{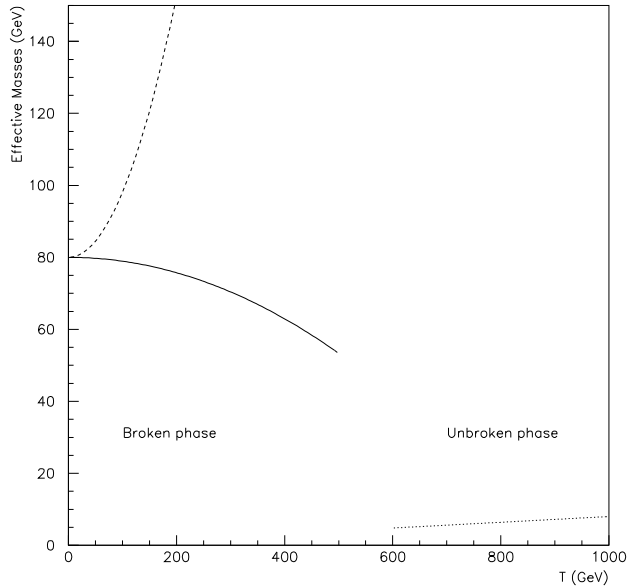}
}
\caption{Electric mass (dashed line) and magnetic mass (solid line) for the 
$W$ gauge boson in the broken phase, and non-perturbative estimate 
of the magnetic mass (dotted line) in the symmetric phase, 
$M_{W,mag}^{sym} = 0.28 g^2 T$.} 
\label{figura}
\end{center}
\end{figure}

These results are shown qualitatively in Fig. 9. 
We plot both the 
magnetic and the electric masses of the $W$ gauge boson as functions 
of the temperature, at leading order $T^2$. The solid and dashed 
lines correspond 
to our one loop calculation, and the dotted line to the 
non-perturbative estimate of the magnetic mass in the 
symmetric phase, $M_{W,mag}^{sym} = 0.28 g^2 T$, 
which is taken from ref. \cite{buch}.

\subsubsection{Model dependence}
\label{model}
As already mentioned, the lowest order term in the 
derivative expansion of the effective Lagrangian, 
${\cal L}_{GChL}$, has a universal character. The next term
in the expansion, ${\cal L}^{(4)}$, is model dependent, 
namely, it depends on the specific dynamics of the symmetry 
breaking sector through the different values of the various constants. 
The logarithmic divergences generated at one loop by 
${\cal L}_{GChL}$ are consistently absorbed by the renormalization 
of those constants. 
As stated before, we have neglected all the zero temperature 
renormalization effects, and therefore the model dependence 
contained on them, as well as in the matching conditions
\footnote{We thank J. Matias for pointing out this fact to us.}.
This is justified, since we are looking for temperature effects.

The only model dependent contribution to the pure thermal  
corrections at one loop is due to the dimension two term
\begin{equation}
{\cal L}_{\beta} = \frac{1}{4} \beta v^2 (Tr (U \tau_3 U^{\dagger} 
(D_{\mu} U) U^{\dagger}))^2 ,
\end{equation}
which explicitly breaks the custodial $SU(2)_C$ symmetry.
This term contributes to $\Delta \rho$ at tree level, and is thus 
strongly constrained by experimental data.  
The contribution of ${\cal L}_{\beta}$ to the magnetic mass of the 
gauge bosons is given by 
\bea
\delta M_{W,mag}^{2} &=& - g^2 \beta \frac{T^2}{12}, \\
\delta M_{Z,mag}^{2} &=& g^2 \beta \frac{T^2}{3}.
\label{surprise}
\eea
The sign of the parameter $\beta$ can be either positive or negative.
Whatever the case for a given theory, eqs. (46) and (47) show an opposite 
behaviour for the $W$ and $Z$ magnetic masses, which are no more forced to
behave alike since the operator under study breaks the custodial $SU(2)_C$ 
symmetry. When combined with the universal leading contribution found 
in eqs. (\ref{mmag}), the total correction
 reads

\bea
M_{W,mag}^{2} &=& \frac{g^2}{4} v^2 (1 - \frac{T^2}{6 v^2} - \beta 
\frac{T^2}{3 v^2}), \\ 
M_{Z,mag}^{2} &=& \frac{(g^2 + g'^2)}{4} v^2 (1 - \frac{T^2}{6 v^2} + \beta
\frac{4 g^2 T^2}{3 (g^2 + g'^2) v^2}).
\label{lasdos}
\eea

 The $SU(2)\otimes U(1)$ symmetry can be considered effectively restored 
only when all possible ``flags'' have signaled it. The above result, taken 
at face value, would indicate that the $SU(2)\otimes U(1)$ symmetry may 
never be restored at high temperature for theories where a large enough 
value of the coefficient of the
 operator ${\cal L}_\beta$  is generated. 
Such a strong statement has to be tempered by recalling that, if the chiral 
expansion is valid, we expect $\beta$ to be small (in typical models it is 
of order of a coupling constant squared), and the total correction in 
(\ref{lasdos}) would be dominated by the leading one, pointing in a 
natural way towards restoration.
Moreover, low energy constraints on new physics give an experimentally 
allowed value of $\beta$ of order $10^{-3}$ \cite{das}, which implies 
that in phenomenologically acceptable models the contribution of 
the operator ${\cal L}_\beta$ is indeed negligible.
It is interesting to retain that the tendency to symmetry restoration 
can be reversed for values 
of $\beta$ which are not outrageously large, as seen from 
eqs. (\ref{lasdos}).

\begin{figure}
\begin{center}
\leavevmode   
\epsfysize=1.8truecm
\epsffile{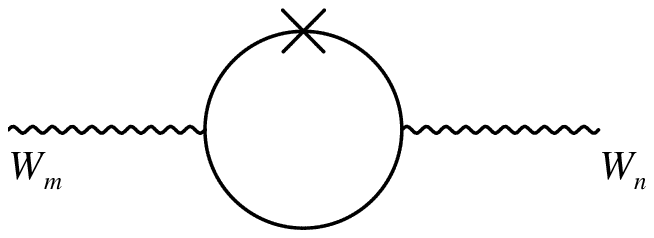}
\epsfysize=1.6truecm
\epsffile{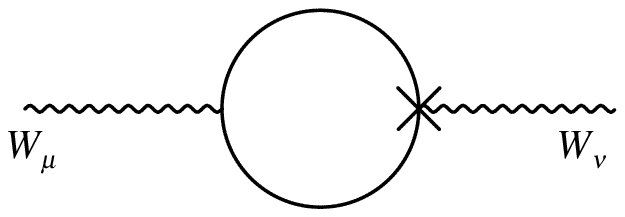}
\epsfysize=1.8truecm
\epsffile{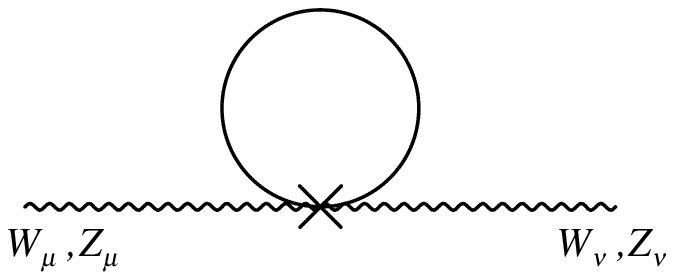}
\\
\caption{Contribution of the model dependent term ${\cal L}_{\beta}$
to the one loop self-energy of the $W$ and $Z$ gauge bosons.}
\label{betagb}
\end{center} \end{figure}

\section{Conclusions}

 We show that the spontaneously broken $SU(2) \otimes U(1)$ gauge theory
 in models where the Higgs sector becomes strongly interacting 
(such as composite Higgs 
 models and  technicolor-like ones) tends to be restored when the system 
is heated. This conclusion is obtained in a model-independent way using 
the techniques of the electroweak chiral Lagrangian. Specific models will
only affect the sharpness of such a tendency, unless the natural 
chiral expansion is not respected. We quantify such model dependence computing
the generic contribution of the leading effective operator whose coefficient
is model-sensitive, ${\cal L}_\beta$; its one-loop contribution to the $W$ 
and $Z$ magnetic masses is found to have opposite sign. 
The technique, while only valid for temperatures
lower than the electroweak scale, has the advantage of its non-perturbative 
character. The physical conclusion reached here parallels the corresponding 
one for the other main avenue of beyond the standard model physics, 
supersymmetry, where perturbative treatments show a tendency towards
restoration. 

 In this work we have as well explored the $SU(2) \otimes U(1)$ gauge 
symmetry in a perturbative regime: 
the cases of one and two light Higgs doublets.
Again, the results show symmetry restoration at high temperatures when 
the full
scalar, gauge boson and fermion corrections are taken into account.

  The above conclusions have been obtained mainly through the study of the 
temperature dependent magnetic mass for the gauge bosons, which we propose 
as an appropriate ``flag'' in the broken phase.

  In addition, $BRST$ invariance has been explicitly
 checked for gauge theories at finite temperature, a novel result.

 Finally, it is worth to remark that global symmetries have been studied as
well for the non-linear sigma model at finite temperature. While this subject 
and the results are not new, the technical approach we used is so: we derive
 first the temperature corrected one-loop effective Lagrangian, from which 
the physical conclusions are then extracted.

 A grain of salt: we have disregarded the putative role of finite 
parameters such as finite volumes  or causal domain sizes in the 
history of the universe. Their effect could constitute an interesting topic 
to study.

\section*{Appendix A: Tensor Basis}

The tensor basis in terms of which we have expressed the gauge boson 
self-energy is given by,
\begin{eqnarray}
A^{\mu \nu} &=& g^{\mu \nu} - B^{\mu \nu} - D^{\mu \nu} \\
B^{\mu \nu} &=& - \frac{\bar{K}^{\mu} \bar{K}^{\nu}}{K^2} \\
C^{\mu \nu} &=& \frac{K^{\mu} \bar{K}^{\nu} + \bar{K}^{\mu} K^{\nu}}{K^2} \\
D^{\mu \nu} &=& \frac{K^{\mu} K^{\nu}}{K^2}
\end{eqnarray}
where $\bar{K}^{\mu} = (K \cdot u \, K^{\mu} - K^2 \, u^{\mu}) / k$ and 
$k$ is such that $K_{\mu} K^{\mu} = \omega^2 - k^2$ with $\omega = 
K_{\mu} u^{\mu}$.

\section*{Appendix B: Faddeev-Popov Lagrangian}

The Faddeev-Popov Lagrangian which corresponds to the non-linear 
realization of the $SU(2) \otimes U(1)$ gauge symmetry is different from 
the one derived for the minimal standard model for which the gauge 
symmetry is linearly realized. Here we present the Faddeev-Popov 
Lagrangian terms which are relevant for our purposes. More general 
results can be found in \cite{Ester}.

\begin{eqnarray}
{\cal L}_{FP} = c_{0}^{+} [- \nabla^2 - (\frac{g'^2 v \xi}{2}) 
[\frac{v}{2} - \frac{1}{6 v} (\pi_{1}^{2} + \pi_{2}^{2}) + ...]] c_{0}
+ \\ \nonumber
\sum_{i \neq j \neq k = 1}^{3} c_{i}^{+} [- \nabla^2 - (\frac{g^2 v 
\xi}{2}) [\frac{v}{2} - \frac{1}{6 v} (\pi_{j}^{2} + \pi_{k}^{2}) + 
...]] c_i + \\ \nonumber
 (c_{1}^{+} c_{2} - c_{2}^{+} c_1) [- g \partial^{\mu} W_{\mu}^{3} +  
(\frac{g^2 v \xi}{2}) \frac{\pi_3}{2}] + ~~~~~~~ \\ \nonumber
 (c_{1}^{+} c_{3} - c_{3}^{+} c_1) [g \partial^{\mu} W_{\mu}^{2} +  
(\frac{g^2 v \xi}{2}) \frac{\pi_2}{2}] + ~~~~~~ \\ \nonumber
 (c_{2}^{+} c_{3} - c_{3}^{+} c_2) [g \partial^{\mu} W_{\mu}^{2} +   
(\frac{g^2 v \xi}{2}) \frac{\pi_1}{2}] + ~~~~~ \\ \nonumber
g g' \frac{v \xi}{4} (c_{0}^{\dagger} c_{3} + c_{3}^{\dagger} 
c_0) [v - \frac{1}{3 v} (\pi_{1}^{2} + \pi_{2}^{2}) + ...] + ...
\end{eqnarray}
where $\xi_1 = \xi_2 = \xi$.

\section*{Appendix C: Ward Identities at One Loop}

We have verified the Ward identities at one loop, both for the linear and 
non-linear realization of the gauge symmetry, for the two-point 
Green functions, computing the $W^{\pm}_{\mu}$, $Z_{\mu}$, $A_{\mu}$ gauge
boson, Goldstone boson and ghost self-energies, together with the
one loop Goldstone boson-gauge boson mixing term. At
leading order (${\cal O} (T^2)$) and for small external momenta 
the results for the standard model case are:

\bigskip
\begin{eqnarray}
\Pi_{D}^{Z} &=& - (g^2 + g'^2) \frac{T^2}{8} - (g^2 + g'^2)
\frac{3 M_Z^2 T^2}{8 m^2}, \nonumber \\
\Pi_{D}^{W} &=& - g^2 \frac{T^2}{8} - g^2 \frac{3 M_W^2 T^2}{8 m^2}, 
\nonumber \\
\Pi_{D}^{\gamma} &=& 0, \nonumber \\
\Pi_{D}^{Z \gamma} &=& 0, \nonumber \\
\Pi^{\pi} &=& 0, \nonumber \\
\Pi^{\pi^{0} \gamma} &=& 0, \nonumber \\
\Pi^{\pi^{\pm} W} &=& - g^2 \frac{T^2}{16 M_W} - g^2 \frac{3 M_W T^2}{16 
m^2}, 
\nonumber \\
\Pi^{\pi^{0} Z} &=&  i (g^2 + g'^2) \frac{T^2}{16 M_Z} + i (g^2 + g'^2)
\frac{3 M_Z T^2}{16 m^2}, 
\end{eqnarray}
\bigskip
where $m^2 = 2 \mu^2$ represents the Higgs boson mass squared.

Concerning the electroweak chiral Lagrangian the results for the two 
point Green functions are:

\begin{eqnarray}
\Pi_{D}^{Z} &=& - (g^2 + g'^2) \frac{T^2}{24} , \nonumber \\
\Pi_{D}^{W} &=& - g^2 \frac{T^2}{24} , \nonumber \\
\Pi_{D}^{\gamma} &=& 0 , \nonumber \\
\Pi_{D}^{Z \gamma} &=& 0 , \nonumber \\
\Pi^{\pi} &=& k^2 \frac{T^2}{18 v^2} , \nonumber \\
\Pi^{\pi \gamma} &=& 0 , \nonumber \\
\Pi^{\pi^{0} Z} &=& i g \frac{T^2}{18 v} , \nonumber \\
\Pi^{c^{\pm}} &=& - \xi g^2 \frac{T^2}{72} .
\end{eqnarray}

\section*{Acknowledgements.}

This work was supported through founds
from CICYT, projects AEN93-0673, AEN96-1718, AEN97-1678, from DGICYT under
grant PB95-1077 and from EEC under the TMR contract ERBFMRX-CT96-0090.
O. P\`ene thanks the BBV fundation for support. 
We are specially indebted to Maria Jos\'e Herrero for inumerable illuminating
discussions. We acknowledge as well Rolf Baier, 
Stefano Catani, Concha Gonz\'alez-Garc\'{\i}a, 
Jean-Pierre Leroy, Cristina Manuel, 
Joaquim Matias, Agust\'{\i}n Nieto, Dominique Schiff 
and Raymond Stora for several inspiring comments.


\begin{thebibliography}{99}
\bibitem{Linde}  D.A. Kirzhnits and A.D. Linde, 
{\it Phys. Lett.} {\bf B42} (1972) 471.
\bibitem{DJ} L. Dolan and R. Jackiw, {\it Phys. Rev.} {\bf D9} (1974) 3320.
\bibitem{Weinberg} S. Weinberg, {\it Phys. Rev.} {\bf D9} (1974) 3357.
\bibitem{fight1} G. Bimonte and G. Lozano, 
{\it Phys. Lett.} {\bf B366} (1996) 248; 
{\it Nucl. Phys.} {\bf B460} (1996) 155.  
\bibitem{fight2} G. Amelino-Camelia, {\it Phys. Lett.} {\bf B388} (1996) 776; 
{\it Nucl. Phys.} {\bf B476} (1996) 255.
\bibitem{fight3} T. Ross, {\it Phys. Rev.} {\bf D54} (1996) 2944; 
M. Pietroni, N. Rius and N. Tetradis, {\it Phys. Lett.} {\bf B397} (1997) 119.
\bibitem{fight4} J. Orloff, {\it Phys. Lett.} {\bf B403} (1997) 309.
\bibitem{fight5} 
Y. Fujimoto and S. Sakakibara, {\it Phys. Lett.} {\bf B151} (1985) 260;
E. Manesis and S. Sakakibara, {\it Phys. Lett.} {\bf B157} (1985) 287; 
G.A. Hajj and P.N. Stevenson, {\it Phys. Rev.} {\bf D37} (1988) 413; 
K.G. Klimenko, {\it Z. Phys.} {\bf C43} (1989) 591; 
{\it Theor. Math. Phys.} {\bf 80} (1989) 929; 
M.P. Grabowski, {\it Z. Phys.} {\bf C48} (1990) 505.
\bibitem{fight6} Y. Fujimoto, A. Wipf and H. Yoneyama, 
{\it Z. Phys.} {\bf C35} (1987) 351; 
{\it Phys. Rev.} {\bf D38} (1988) 2625.
\bibitem{fight7} C. King and L.G. Yaffe, 
{\it Comm. Math. Phys.} {\bf 108} (1987) 423;
G. Bimonte and G. Lozano, {\it Phys. Lett.} {\bf B388} (1996) 692. 
\bibitem{fight8} G. Bimonte et al., DFTUZ 97/17, hep-lat/9707029,
to appear in {\it Nucl. Phys.} {\bf B}.
\bibitem{axion} R.N. Mohapatra and G. Senjanovic, 
{\it Phys. Lett.} {\bf B89} (1979) 57; 
{\it Phys. Rev. Lett.} {\bf 42} (1979) 1651; 
{\it Phys. Rev.} {\bf D20} (1979) 3390;
G. Dvali, G. Senjanovic, {\it Phys. Rev. Lett.} {\bf 74} (1995) 5178;
G. Dvali, A. Melfo, G. Senjanovic, {\it Phys. Rev.} {\bf D54} (1996) 7857.
\bibitem{monopole} G. Dvali, A. Melfo, G. Senjanovic,
{\it Phys. Rev. Lett.} {\bf 75} (1995) 4559.
\bibitem{haber} H. Haber, {\it Phys. Rev.} {\bf D46} (1982) 1317; 
M. Mangano, {\it Phys. Lett.} {\bf B147} (1984) 307.
\bibitem{central} A. Riotto and G. Senjanovic, 
{\it Phys. Rev. Lett.} {\bf 79} (1997) 349.
\bibitem{senja} G. Dvali, K. Tamvakis, {\it Phys. Lett.} {\bf B378} (1996) 141;
B. Bajc, A. Melfo, G. Senjanovic, {\it Phys. Lett.} {\bf B387} (1996) 796.
\bibitem{stern} J. Stern,  How to see the chiral structure of qcd 
vacuum in low-energy $\pi-\pi$ scattering. e-print archive: 
hep-ph/9510318. Contribution to the 2nd Workshop on Physics and Detectors 
for DAPHNE 95.
\bibitem{elitzur} S. Elitzur, {\it Phys. Rev.} {\bf D12} (1975) 3978.
\bibitem{buch} W. Buchm\"uller and O. Philipsen, 
{\it Phys. Lett.} {\bf B397} (1997) 112. 
\bibitem{ptyt} R.D. Pisarski and M. Tytgat, 
{\it Phys. Rev.} {\bf D54} (1996) 2989, and references therein.
\bibitem{Kapusta}  A. Bochkarev, J. Kapusta, 
{\it Phys. Rev.} {\bf D54} (1996) 4066.
\bibitem{Leut1} J. Gasser, H. Leutwyler, {\it Phys. Lett.} {\bf B184} 
(1987) 83.
\bibitem{Schenk} A. Schenk, {\it Phys. Rev.} {\bf D47} (1993) 5138.
\bibitem{Gaillard} P. Binetruy and M.K. Gaillard, {\it Phys. Rev.} {\bf D32}
(1985) 931. 
\bibitem{ptyt2} R.D. Pisarski and M. Tytgat, 
{\it Phys. Rev. Lett.} {\bf 78} (1997) 3622.
\bibitem{Cris} C. Manuel, ECM-UB-PF-97/27, hep-ph/9710208, to appear 
in {\it Phys. Rev.} {\bf D}. 
\bibitem{toublan} D. Toublan, 
{\it Phys. Rev.} {\bf D56} (1997) 5629; 
\bibitem{basis} M. Le Bellac, {\it Thermal Field Theory},
Cambridge University Press, Cambridge 1996. 
\bibitem{al} T. Appelquist and C. Bernard, 
{\it Phys. Rev.} {\bf D22} (1980) 200; 
A.C. Longhitano, {\it Phys. Rev.} {\bf D22} (1980) 1166;
{\it Nucl. Phys.} {\bf B188} (1981) 118.
\bibitem{bp} R.D. Pisarski, {\it Phys. Rev. Lett.} {\bf 63} (1989) 1129; 
E. Braaten and R.D. Pisarski, {\it Nucl. Phys.} {\bf B337} (1990) 569.
\bibitem{mag} A.D. Linde, {\it Phys. Lett.} {\bf B96} (1980) 289;
D. Gross, R. Pisarski and L. Yaffe, {\it Rev. Mod. Phys.} 
{\bf 53} (1981) 43.  
\bibitem{das} S. Dawson, S. Alam and R. Szalapski, 
{\it Phys. Rev.} {\bf D57} (1998) 1577.
\bibitem{Ester} M.J. Herrero and E. Ruiz-Morales, {\it Nucl. Phys.} 
{\bf B418} (1994) 431.
\bibitem{Quim} D. Espriu and J. Matias, {\it phys. Rev.} 
{\bf D52} (1995) 6530. 
\end{thebibliography}
\end{document}